# Performance Analysis of 2*4 MIMO-MC-CDMA in Rayleigh Fading Channel Using ZF-decoder

Mr. Atul Singh Kushwah

*Asst. Professor & Digital Communication & Indore Institute of Science & Technology-II, Indore (M.P), India*
atulsingh41189@gmail.com

*Abstract* - In this paper we analyze the performance of 2*4 MIMO-MC-CDMA system in MATLAB which highly reduces BER. In this paper we combine MIMO and MC-CDMA system to reduce bit error rate in which MC-CDMA is multi user and multiple access schemes which is used to increase the data rate of the system. MC-CDMA system is a single wideband frequency selective carrier which converts frequency selective to parallel narrowband flat fading multiple sub-carriers to enhance the performance of system. Now MC-CDMA system further improved by grouping with 2*4 MIMO system which uses ZF (Zero Forcing) decoder at the receiver to decrease BER with ½ rate convolutionally encoded Alamouti STBC block code is used as transmit diversity of MIMO through multiple transmit antenna. Importance of using MIMO-MC-CDMA using convolution code is firstly to reduce the complexity of system secondary to reduce BER and lastly to increase gain. In this paper we examine system performance in diverse modulation techniques like, 8-PSK, 16-QAM, QPSK, 32-QAM, 8-QAM and 64-QAM in Rayleigh fading channel using MATLAB.

I.                Introduction

Due to advancement in technology high data rate and low probability of error are today's requirement in this paper we implement CDMA, OFDM and MIMO, results advanced technique for reducing error rate. MC-CDMA is multiple access and multicarrier system which forms by the combination of OFDM and CDMA [11]. This MC-CDMA is frequency selective flat fading channel. The MC-CDMA improved efficiency of the wireless communication system thereby resulting high data rate and low probability of error.

In this paper MC-CDMA is combined with MIMO to increase the performance. 2*4 MIMO is multiple antenna system in which two transmit antenna and four receive antenna are used and for detection multiple receive diversity and multiple transmit diversity with half-rate convolutionally encoded Alamouti STBC code is used and also for the synchronization of system to reduce Inter Symbol Interference. To detect orthogonality of signal Zero-Forcing(ZF) detection scheme is used. And finally MIMO-MC-CDMA [4] is formed by above operations in MATLAB then it will be analyzed in different modulation techniques like 8-PSK, 16-QAM, QPSK, 32-QAM, 8-QAM and 64-QAM in Rayleigh fading channel.

II.    MULTIPLE INPUT MULTIPLE OUTPUT (MIMO)

In this paper we use MIMO systems [2] with two transmit antennas and four receive antennas at the receiver, as a result both transmit and receive diversity techniques are applied to decrease fading resulting from signal distortion by channel. Independently faded multiple data are reproduced at the receiver which is repetitive in nature that is replicas are received at the receiver. The system gives the gain in diversity which shows the difference in SNR at the output of the diversity combiner as compared to that of single diversity at definite probability level.

A MIMO system consisting of N number of transmitting antenna elements equal to two, and of M number of receive antenna elements equal to four was modeled, that is diversity order of 4 can be achieved. This paper utilizes Zero Forcing (ZF) decoder to combine M number of received signals to resonate on the frequently required desired transmitted signals. The sum of the received SNR from 4 various paths is the effectively received SNR of the system with Alamouti's STBC with 2*4 diversity. The number of receivers necessary to demodulate all 4 received signals in case of ZF decoder for a source with independent signals in the received antennas.

### III. MULTI CARRIER CODE DIVISION MULTIPLE ACCESS (MC-CDMA)

OFDM and CDMA [5] combination forms the enhanced system called MC-CDMA [3,1,6]. In this system spreading of data is done using PN sequence and modulation is done to input data signal. MC-CDMA combines the advantage of multipath fading of OFDM system with the advantage of multi-user access of CDMA system.

*A. Mathematical model*

In present MC-CDMA system we consider two transmitting and four receiving antenna for signal detection scheme. Assume H refers to channel matrix as represented by $h_{ij}$ for the channel gain between the i-th and j-th i.e. transmitting antenna and receiving antenna respectively, j=1,2,3,4 and i=1,2. Each user data is represented by $a=[a_1, a_2]^T$ and the equivalent received signals are represented by $y=[y_1\ y_2\ y_3\ y_4]^T$, in which $a_i$ denote the transmit signal from i-th transmitting antenna and $y_j$ denote the received signal at the j-th received antenna. Suppose $n_j$ denote the Additive white Gaussian noise of variance $\sigma_{n4}$ at the receiving antenna j-th and $h_i$ shows the i-th column vector of channel matrix H. Then received signal y for system can be given by

$$y=Ha+n \quad \ldots\ldots\ldots\ldots(1)$$

Where, $n=[n_1,n_2,n_3,n_4]^T$

As the interfering signals by other transmitting antennas is decreased for detecting the preferred signal from the transmitting antenna, the detected preferred signal from the transmitting antennas by inverting channel effect by a weight matrix W is

$$\tilde{A}=[\tilde{a}_1\ \tilde{a}_2]=W_y\ldots\ldots\ldots(2)$$

For Zero-Forcing (ZF) detection scheme, the weight of matrix is given by

$$W_{ZF}=(H^H H)^{-1} H^H \quad \ldots\ldots(3)$$

where $()^H$ shows the Hermitian transpose and the detected preferred signal from the transmitting antenna using the following

$$\tilde{a}_{ZF}=W_{ZF}y \quad \ldots\ldots\ldots(4)$$

It is anticipated to find transmitted signal vector by Sphere Decoding (SD) scheme by minimum least ML metric. Assume $y_R$ and $y_I$ shows the real and imaginary parts of received signal y, i.e. $y_R$= Re{y} and $y_I$= Im{y}. Analogous to input signal $x_i$ and the channel gain $h_{ij}$ from ith transmitting antenna toward jth receiving antenna be able to be shown by $a_{iR}$ = Re{$a_i$} and $a_{iI}$ = Im{$a_i$} and $h_{ijR}$ = Re{$h_{ij}$} and $h_{ijI}$ = Im{$h_{ij}$} correspondingly. Mathematically,

$$\hat{y}=[y_R\ jy_I]^T \ldots\ldots\ldots(5)$$

From (6), the detected preferred signal $\tilde{a}_{SP}$ with real and imaginary mechanism from the transmitting antenna be able to be shown by [3]:

$$\tilde{a}_{SP}=[\tilde{a}_{1R}\ \tilde{a}_{2R}\ \tilde{a}_{1I}\ \tilde{a}_{2I}]^T =[\hat{H}^T\ \hat{H}]^{-1}*[\hat{H}^T\ \hat{y}]\ldots(6)$$

### IV. MIMO-MC-CDMA COMMUNICATION SYSTEM MODEL

Communication system model of MIMO-MC-CDMA is shown in fig.1.

In MIMO-MC-CDMA communication system we are assume transmitter sends arbitrary sequence to the receiver so we used random PN sequence generator by MATLAB. Now spreading of sequence is done by PN sequence generator. Then in modulator different modulation schemes are used like *8-QAM, QPSK, 8-PSK, 16-QAM, 32-QAM and 64-QAM* this is shown by modulator block. MC-CDMA system which is already described in section 3 by Multi-Carrier Code Division Multiple Access (MC-CDMA). Now MIMO (multiple Input Multiple Output) encoder ½ rate convolutionally encoded Alamouti's STBC block code is used which will be explained in section 3 as Multiple Input Multiple Output (MIMO). Combination of MC-CDMA and MIMO form MIMO-MC-CDMA as shown in fig.1. Now signal is transmitted by Rayleigh Fading Channel [7]. Now receiver receive the signal in reverse manner with ZF decoder for the regeneration of transmitted signal at receiver and BER calculation is done for determining the system performance. In MIMO two transmit and four receive antennas are used. In this paper we are transmitting message bits which is random in nature or else dependent on user then this data is passed through spreader using PN sequence which forms 8 bits for each input bits that is input bits*8 then ensuing bits are formed after the spreading of encoded sequence. Then these spreaded sequence of bits are passed by modulator in which its modulation depends on the type of modulation used. Now these modulated data is reframed into parallel form for OFDM then IFFT is done to convert frequency selective wide-band carriers into parallel narrowband flat-fading carriers which are orthogonal in character then this data is reshaped to parallel to serial then CP cyclic prefix addition is done to remove ISI which complete the process of OFDM transmission then this data then passed through MIMO encoder which uses Alamouti STBC code for 2 transmit and 4 receive diversity techniques in which 4*2 channel matrix is created by using MIMO diversity, and ZF detection scheme is used at the receiver then reverse operation is done for receiving the input bits.

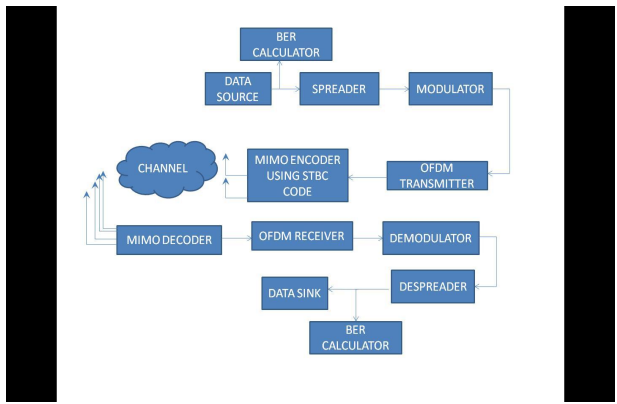

Fig.1. Communication System Model OF 2*1 MIMO-MC-CDMA

V. SIMULATION RESULTS AND DISCUSSION:

Table 1 shows the input model parameters of MIMO-MC-CDMA [7,8,9,10] in different modulation technique.

Fig.2 shows the comparative analysis of different modulation schemes in MIMO-MC-CDMA.

Table 2 represents the performance analysis of various modulation schemes in terms of gain and BER.

TABLE I

**SIMULATED MODEL PARAMETERS**

| Channel Encoder | ½ rate convolution encoder Alamouti STBC |
|---|---|
| Signal detection scheme | Zero forcing |
| Channel | Rayleigh Fading Channel |
| Signal to Noise Ratio | -10dB to 20 dB |
| CP Length | 1280 |
| OFDM Sub-carriers | 6400 |
| No. of transmitting and receiving antennas | 2*4 |
| Modulation Schemes | QPSK, 8-PSK, 8-QAM, 16-QAM, 32-QAM and 64 QAM |

From table.2 and Fig.2 we can say that QPSK shows high gain (12.23 dB) with low BER with respect to other modulation schemes at -5dB SNR. This is done by using MIMO-MC-CDMA system in which error probability in QPSK is zero which shows low probability of error in system.

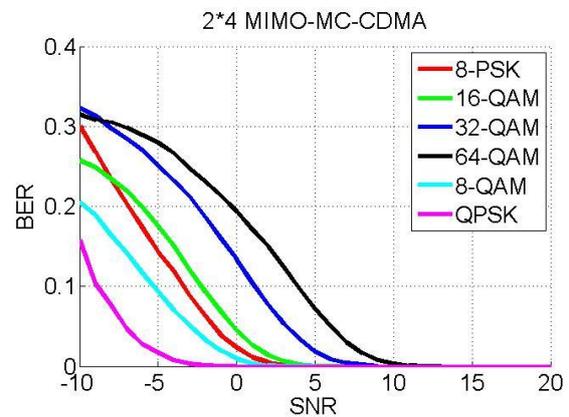

Fig.2. Performance analysis of MIMO-MC-CDMA in 8-QAM, 16-QAM, 32-QAM, 64-QAM, 8-PSK and QPSK modulation scheme.

TABLE II

**PERFORMANCE ANALYSIS AT -5DB SNR WITH RESPECT TO 64-QAM MODULATION TECHNIQUE AS SHOWN IN FIG.4**

| Modulation | BER at -5dB | Gain w.r.t 64-QAM |
|---|---|---|
| QPSK | 0.01673 | 12.23 dB |
| 8-QAM | 0.09321 | 9.4 dB |
| 8-PSK | 0.1428 | 2.918 dB |
| 16-QAM | 0.1754 | 2.025 dB |
| 32-QAM | 0.2511 | 0.4669 dB |
| 64-QAM | 0.2796 | 0dB |

## VI. CONCLUSION

Fig.2 represents the comparative analysis of MIMO-MC-CDMA in different modulation schemes. Table 2 represents the comparative analysis for different modulation schemes which shows that as modulation order is higher results increase in BER. This manuscript aims to decrease bit error rate which is found by QPSK modulation scheme with gain of 12.23 dB with respect to 64-QAM which shows that the gain of QPSK is higher in comparison to other modulation technique with low probability of error because errors were finished at 0dB in QPSK. For 3G and 4G communication 64-QAM modulation scheme is preferred which contain BER up to 10dB, i.e. errors are removed in 64-QAM at 10dB SNR which is optimized by using MIMO-MC-CDMA.